# How Market Structure Drives Commodity Prices

**Bin Li[1]†, K. Y. Michael Wong[1]\*, Amos H. M. Chan[1]‡, Tsz Yan So[1]§,**

**Hermanni Heimonen[1]‖, Junyi Wei[1], and David Saad[2]**

[1]Department of Physics, The Hong Kong University of Science and Technology,

Clear Water Bay, Hong Kong.

[2]The Nonlinearity and Complexity Research Group, Aston University,
Birmingham B4 7ET, United Kingdom.

## Abstract

We introduce an agent-based model, in which agents set their prices to maximize profit. At steady state the market self-organizes into three groups: excess producers, consumers and balanced agents, with prices determined by their own resource level and a couple of macroscopic parameters that emerge naturally from the analysis, akin to mean-field parameters in statistical mechanics. When resources are scarce prices rise sharply below a turning point that marks the disappearance of excess producers. To compare the model with real empirical data, we study the relations between commodity prices and stock-to-use ratios of a range of commodities such as agricultural products and metals. By introducing an elasticity parameter to mitigate noise and long-term changes in commodities data, we confirm the trend of rising prices, provide evidence for turning points, and indicate yield points for less essential commodities.

PACS 89.65.Gh 05.20.-y 89.75.-k

Keywords: market structure; commodity prices; agent-based model; inventory; turning point; stock-to-use ratio

\* E-mail address: phkywong@ust.hk
† Present address: Shenzhen Middle School, 18 Shenzhong Street, Renmin North Road, Luohu District, Shenzhen, Guangdong, China.
‡ Present address: Theoretical Physics, University of Oxford, Keble Road, Oxford OX1 3NP, United Kingdom.
§ Present address: Room 6.10, 6/F, The Jockey Club Tower, Centennial Campus, The University of Hong Kong, Pokfulam Road, Hong Kong.
‖ Present address: Centre for Quantum Technologies, National University of Singapore, 3 Science Drive 2, 117543, Singapore.

**Introduction**

The oil price crisis of 1973 rattled the world and left persistent effects on the world economy and politics [1]. Peak periods in food price index during 2008 and 2011 coincided with incidents of food riots and instabilities across the world [2]. Clearly, prices of commodities affect our lives in many ways; they determine the economic well-being of individuals, companies, societies and the stability of governments. Besides the immediate effects on the livelihood of the average citizen, farmers need to know the prices of crops for planning their land use, manufacturers need to know when to import their raw materials, policy-makers need to decide on their agricultural stabilization schemes, and speculators would like to make a fortune in the futures market. Correlations exist between stock level and prices [3], and were illustrated by the 2008 hike in grain price due to the diversion of corn to biofuel production [4]. When stocks decline to dangerous levels, prices become highly sensitive to small perturbations.

Since factors affecting commodity prices are complex, most analyses are based on empirical fitting, such as power-law relations between price and stock level [5]. However, few models are based on fundamental principles that relate prices to the microscopic composition of markets [6]. The effect of changes in market structure on prices was illustrated by the famous Lewis Model describing the labor market in developing economies, in which wages rise due to shortage in low-cost labor [7], as has been experienced in China recently [8].

In this work we devise an agent-based model to better understand how market structure drives prices. It has the advantage of capturing the most essential elements of market structure that affects commodity prices. Its amenability to quantitative analysis enables us to vary the macroscopic parameters of the market and identify different regimes of market structure. The composition of the agent population in each regime is different, yielding regimes of inelastic and elastic price behaviors. This indicates that the empirical fitting of price trends in different price regimes by a single function is not sufficient.

To verify the prediction of our model, we further study empirical data of commodity stocks and prices. Although such data trends are often obscured by noises, we significantly improve the quality of the data using regressions of an elasticity parameter. As a result, we are able to reveal the existence of turning points from inelastic to elastic regimes in some commodities in agreement with our model. Furthermore, we discover from the data the existence of yield regimes in some commodities.

In a nutshell, the success of our model is due to the inclusion of the following components. First, it identifies the uneven distribution of resource as an important component of the model and a major driver of price changes. This determines the market structure composed of agents with different capacities playing different roles in the market as excess producers, balanced agents or consumers. They set prices differently in the market since microscopically the bargaining power of the agents depends on supply and demand. In a sufficiently well-connected market sellers with more abundant supply may set lower prices so as to capture a larger market share and buyers with a strong demand may accommodate higher prices to secure commodity provision.

The second component of the model is market interaction. With the recent application of social network theory to economics [9], the bargaining power of the agents was found to depend on the topology of the corresponding trading networks, which determines the competition relation between suppliers and consumers [10], giving rise to price variations at equilibrium [11].

Furthermore, such interactions can lead to Nash equilibrium states that maximize the utility of agents [12]. This process of attaining a global stationary state through local responses to neighboring interactions can be considered a graphical game [13], implemented through passing messages between neighbors, like those in network optimization and inference [14].

The formalism of our model can be applied to networks with different topologies but in this paper we focus on fully connected networks, representing markets in which agents can freely trade among themselves. In this setting, the many-body effects of the agents' own pricing decisions are diluted, enabling us to describe the market state by only a few macroscopic parameters, resembling mean-field approaches in statistical physics. We can then monitor changes in price trends in different regimes and see whether price changes resemble phase transitions in many-body interacting systems [15].

The third component of the model is the role played by inventory in shaping the transition between the different regimes. Inventory increases the demand of the agents in a soft way that smooths out sharp transitions of prices, enabling our model to reveal the elastic price regime that would be otherwise masked by neglecting inventory effects. Here inventory carries a different meaning from stocks. Stocks refer to the excess amount of commodities left behind in the hands of the agents when their production plus inflow exceeds outflow. On the other hand, low levels of inventory are necessary for all agents to maintain a smooth operation of the system [16]. For example, industrialists need to keep an inventory of raw materials so as to streamline their manufacturing process. Dealers need to keep an inventory to facilitate sales and deliveries to anticipate sporadic transactions, and occasionally they are forced to carry inventories when faced with low seasons of sale. While inventory levels are low, they act as buffers to smooth sharp changes in supply and demand, and represent the level of commodity agents keep to avoid running out of stock when purchasing orders arrive.

**Results**

**Model**

We consider a network of $N$ nodes. Each node $i$ is connected to a set of trading partners denoted as $\partial i$. Unless stated otherwise, we will consider fully connected networks in this work where $\partial i$ consists of all nodes except $i$. Each node is either a producer or a consumer of a commodity, with an initial capacity $\Lambda_i$ randomly drawn from a distribution $\rho(\Lambda_i)$ for node $i = 1, \ldots, N$. Positive $\Lambda_i$ represents the amount of commodity produced per unit time by node $i$, whereas negative $\Lambda_i$ represents the amount of commodity consumed per unit time by node $i$. The commodity is essential to all consumers, so that each consumer has to purchase a sufficient amount of commodity to satisfy their needs, and each producer cannot sell more commodity than its capacity. This is possible globally if the average $\langle \Lambda \rangle$ of the distribution $\rho(\Lambda)$ is positive. Let $y_{ij}$ be the flow of commodity from node $j$ to $i$. We adopt the convention that negative $y_{ij}$ means a flow of magnitude $|y_{ij}|$ in the opposite direction. Hence the inequality $\Sigma_{j \in \partial i} y_{ij} + \Lambda_i \geq 0$ applies to each node $i$. The flows $y_{ij}$ associated with a producer (consumer) $i$ with a largely positive (negative) capacity are all outgoing (incoming), while the flows associated with a node with intermediate capacity may be partly outgoing and partly incoming, corresponding to their role as middle-men besides providing or consuming their own resources.

The net demand raised by node $i$ to other nodes is the outflow minus the capacity if the difference is positive and 0 otherwise, given by $\max(\Sigma_{j \in \partial i} y_{ji} - \Lambda_i, 0)$. When the argument $\Sigma_{j \in \partial i} y_{ji}$

− $\Lambda_i$ changes sign, the demand has a discontinuous slope. In practice, trading nodes need to keep a provisional level of commodity so that they do not run out of stock when purchasing order arrives. Hence we propose a smoother demand $\xi_i$

$$\xi_i = f\left(\sum_{j\in\partial i} y_{ji} - \Lambda_i\right), \tag{1}$$

where $f(x)$ is an inventory-dependent function with a continuous slope, and asymptotically approaches 0 and $x$, respectively, in the limits $x \to \mp\infty$. For convenience, we use $f(x) = v \ln[1+\exp(x/v)]$, where $v$ is referred to as the inventory level, but other functions may also be considered. The original demand function with a discontinuous slope at zero demand is recovered in the limit $v\to 0$. On the other hand, for finite values of $v$, $f(x)$ starts to deviate smoothly from 0 when $x$ is of the same order as $v$. The inventory has the same effect as a fluctuating capacity $\Lambda_i + z_i$, where $z_i$ is drawn from the distribution $P(z_i) = \text{sech}^2(z_i/4v)/4v$.

To satisfy the demand $\xi_i$, node $i$ purchases commodity from other nodes. Let $r_{ij}$ be the fraction purchased from node $j$ by node $i$, so that the amount of commodity shipped from $j$ to $i$ is $y_{ij} = \xi_i r_{ij}$. The fractions are determined by the prices set by neighboring nodes $k \in \partial i$ on a competitive basis. We consider fractions of the form

$$r_{ij} = \frac{F(\phi_j)}{\sum_{k\in\partial i} F(\phi_k)}, \tag{2}$$

where $\phi_j$ is the price set by node $j$, and $F(\phi)$ is a non-negative decreasing function of $\phi$. For convenience, we use the exponential form $F(\phi) = \exp(-\beta\phi)$, where $\beta$ is a parameter playing the role of inverse temperature in the statistical physics literature, but other forms are also possible. When $\beta \to \infty$, $r_{ij}$ becomes a winner-take-all function, such that the node with the lowest price becomes the sole provider of node $i$. In reality, agents diversify their purchases due to many factors. For example, they may have considerations other than prices such as quality and service, they may not like to be monopolized, or the cheapest choice may not be available at their moment of need. We note that $\beta^{-1}$ is the scale of the price. This means that when the prices set by two suppliers differ by less than $\beta^{-1}$, the buyer would purchase from both suppliers with roughly equal weight. However, when the price difference becomes much greater the purchasing amount will differ significantly. Hence $\beta^{-1}$ can be considered as the intrinsic value of a unit of commodity. For convenience, we will take $\beta = 1$, so that prices are scaled in units of the intrinsic value.

Each node $i$ calculates its price $\phi_i$ by minimizing its net cost $E_i$, which is the purchasing cost minus the sales revenue, assuming that the price of other nodes are not changed. Hence

$$E_i = \sum_{j\in\partial i} y_{ij}\phi_j - \sum_{j\in\partial i} y_{ji}\phi_i = \sum_{j\in\partial i} \xi_i r_{ij}\phi_j - \sum_{j\in\partial i} \xi_j r_{ji}\phi_i. \tag{3}$$

The clearing and price adjustment process of this trading model with and without inventory can be simulated in the way described in the Supporting Materials (SM).

**Predictions**

To minimize $E_i$, node (trader) $i$ needs to assess the effects of changing its price by $\delta\phi_i$. Obviously, the sales revenue changes since the price of every unit of sold commodity changes. In addition, node $i$ needs to know how its trading partners respond to the price change, specifically the flow change $\delta y_{ji}$ in response to $\delta\phi_i$. It may obtain this knowledge through an active bargaining process, or through the passive observation of how the sales volume changes with price. Node $i$ will then consider such messages from all neighbors before establishing its new price. In this respect, this trading network model belongs to the class of network problems solvable by passing messages [14]. The message sent from node $j$ to $i$ through the bargaining process is

$$a_{ij} = -\frac{\partial y_{ji}}{\partial \phi_i} = \xi_j r_{ji}(1-r_{ji}) - r_{ji}\frac{\partial \xi_j}{\partial \phi_i}. \tag{4}$$

Since the export of node $i$ changes, the demand $\xi_i$ in the purchasing cost also changes. Using Eq. (1), we have

$$\frac{\partial \xi_i}{\partial \phi_i} = -f'\left(\sum_{j\in\partial i}\xi_j r_{ji} - \Lambda_i\right)\sum_{j\in\partial i} a_{ij}, \tag{5}$$

For the term $\partial\xi_j/\partial\phi_i$ in Eq. (4), we need to consider how a price change $\delta\phi_i$ at node $i$ induces changes in demands of all nodes, assuming that prices at other nodes are unchanged. However, the demand changes are interdependent. $\delta\xi_i$ induces changes in the neighbors of $j$, which induces changes back in $\delta\xi_j$, commonly referred to as Onsager reactions in many-body physics. As shown in the SM for fully connected networks using Green's function techniques, $\delta\xi_j$ is of the order $N^{-1}$ of $\delta\xi_i$ for nodes $j$ neighboring node $i$. Hence the second term in Eq. (4) can be neglected in the large $N$ limit. After collecting messages from all neighbors, the price becomes

$$\phi_i = \frac{\sum_{j\in\partial i}\xi_j r_{ji}}{\sum_{j\in\partial i}\xi_j r_{ji}(1-r_{ji})} + f'\left(\sum_{j\in\partial i}\xi_j r_{ji} - \Lambda_i\right)\sum_{j\in\partial i} r_{ij}\phi_j. \tag{6}$$

When the network is fully connected, the price behavior depends on only two macroscopic parameters: $\phi_p \equiv \langle\phi e^{-\phi}\rangle/\langle e^{-\phi}\rangle$ being the average purchasing price and $y \equiv \langle\xi\rangle/\langle e^{-\phi}\rangle$ termed the demand coefficient, where the outflow from a node to satisfy the network's demand is $ye^{-\phi}$. Averages denoted by the angled brackets are taken over all nodes. The price $\phi$ of a node becomes a unique function of its capacity $\Lambda$, bounded between the maximum price $1+\phi_p$ and minimum price 1. $\phi(\Lambda)$ is the inverse function of

$$\Lambda(\phi) = ye^{-\phi} + v\ln\left(\frac{1+\phi_p - \phi}{\phi - 1}\right). \tag{7}$$

We first consider the limit of zero inventory. When $v \to 0$ the price at node $i$ depends on its capacity $\Lambda_i$ as

$$\phi_i = \begin{cases} 1+\phi_p, & \Lambda_i \leq ye^{-1-\phi_p}, \\ \ln\left(\dfrac{y}{\Lambda_i}\right), & ye^{-1-\phi_p} \leq \Lambda_i \leq ye^{-1}, \\ 1, & ye^{-1} \leq \Lambda_i. \end{cases} \tag{8}$$

Hence there are three types of nodes. (1) Consumers ($\Lambda_i < ye^{-1-\phi_p}$) with positive demands and price $1+\phi_p$. (2) Balanced agents ($ye^{-1-\phi_p} \leq \Lambda_i \leq ye^{-1}$) with zero demands and no excess resources. Their prices are $\ln(y/\Lambda_i)$. (3) Excess producers ($\Lambda_i > ye^{-1}$) with zero demands and excess resources and price 1.

With the inventory effect, price becomes a continuously changing function, as shown in Fig. 1(A). The three groups of nodes can still be identified, although the boundaries become fuzzy. Due to the presence of inventory, the outflows of the balanced nodes differ from their capacities by an amount of order $v$. For balanced agents with $ye^{-1-\phi_p} \leq \Lambda_i \leq ye^{-1-\phi_p/2}$, the outflow is greater than the capacity by an amount of order $v$, whereas for balanced agents with $ye^{-1-\phi_p/2} \leq \Lambda_i \leq ye^{-1}$, the outflow is less than the capacity by an amount of the order $v$. These two groups will be referred to as quasi consumers and quasi producers respectively.

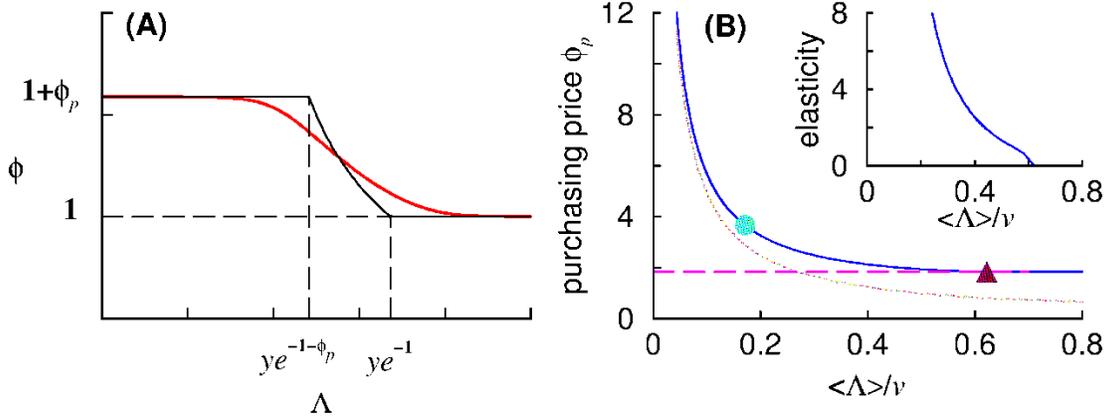

**Fig. 1. Price predictions of the model.** (A) The dependence of price on the node capacity. Red curve: Eq. (7) for $\langle\Lambda\rangle = 0.2$ and $v = 0.01$. Black curve: $v = 0$ limit. (B) The dependence of purchasing price on average capacity (solid blue line). Pink dashed line: price in the regime with excess producers. Brown symbol: Disappearance of excess producers. Turquoise dot: Disappearance of quasi producers. Brown dotted line: asymptotic limit of vanishing capacity. Inset: The dependence of the capacity elasticity of price on the average capacity in the $v = 0$ limit.

Solutions of the self-consistent equations for $\phi_p$ and $y$ depend on the resource distribution $\rho(\Lambda)$. Considering the bounded resource production and consumption in real data we adopt distributions with upper and lower bounds. The expressions of $\phi_p$ and $y$ in the limit of small $v$ are derived in the SM for the rectangular distribution of mean $\langle\Lambda\rangle$ and width 1.

For the rectangular capacity distribution with $v = 0$, the dependence of price and cost on capacity is verified by simulations shown in Figs. 2(A) and (B) respectively. In both figures, the theoretical results (dashed lines) are in excellent agreement with those obtained by solving the Nash equilibrium equations (6). As expected, the cost increases with decreasing capacity. It is interesting to note that through trading at an optimal price, even the consumers with $\Lambda_i$ close to 0 can gain profit (negative cost).

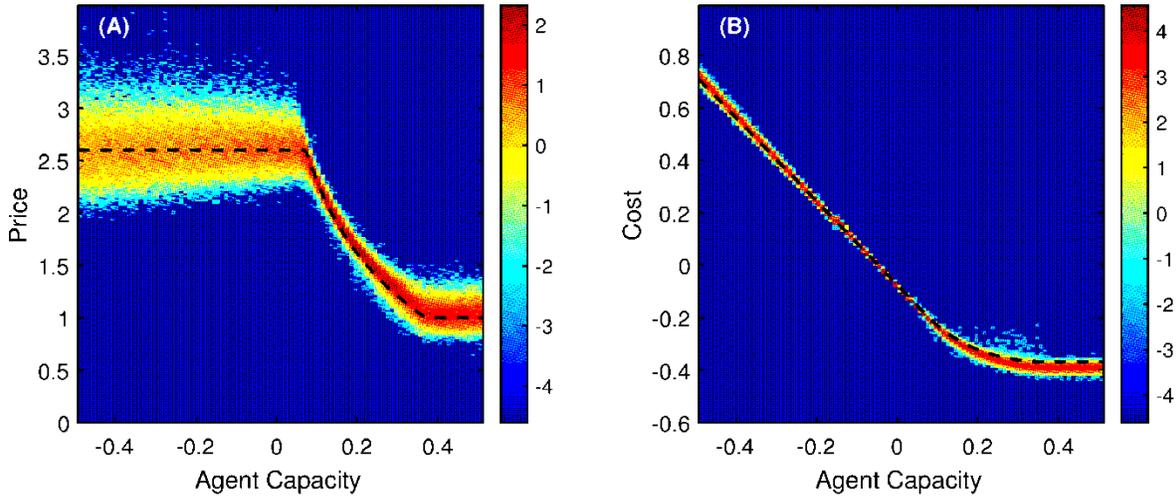

**Fig. 2. Model verification by simulations.** The price distribution (A) and the cost distribution (B) for different agent capacities with $\langle\Lambda\rangle = 0.01$, $v = 0$, $N = 100$, 1,000 samples, 10,000 time steps and 200 updating cycles per time step. Dashed curves: predictions by Eq. (7). Both distributions are in natural log scale.

When resources become increasingly tight, the purchasing price increases, as shown in Fig. S2 of SM. For the rectangular capacity distribution, $\phi_p$ approaches the finite value of 1.83 with an infinite slope when $\langle\Lambda\rangle$ approaches 0. When $\langle\Lambda\rangle$ falls below 0, the price diverges discontinuously. Note that excess producers exist in the range $(\sqrt{1/2}-\sqrt{\langle\Lambda\rangle})^2 \leq \Lambda \leq 1/2 + \langle\Lambda\rangle$, showing that the fraction of excess producers approaches 0 when $\langle\Lambda\rangle$ approaches 0. However, for finite values of $\langle\Lambda\rangle$, excess producers always exist in the case of zero inventory $v = 0$.

When $v$ has a small non-zero value, the price discontinuity for $v = 0$ is smeared out to give a more refined picture in the range $\langle\Lambda\rangle \sim v$ as shown in Fig. 1(B). First, we find that when $\langle\Lambda\rangle$ is in the range $0.622v \leq \langle\Lambda\rangle \ll 1$, the average price remains effectively at 1.83. This is the inelastic regime, and the market structure consists of excess producers, balanced agents and consumers.

When $\langle\Lambda\rangle$ falls below $0.622v$, excess producers disappear, and the price rises above 1.83 (brown symbol in Fig. 1(B)). This is the elastic regime and the market structure consists of only the balanced agents and the consumers. The rise in price shows that in this regime, the excess producers can no longer play the role of stabilizing the price by acting as a reservoir of resources. However, although resource production is still above consumption for $\langle \Lambda \rangle > 0$, the holding up of resources in inventories causes the excess resources of the excess producers to dry up. The price thus experiences a *sharp turning point*. This turning point resembles a phase transition in many physical systems. Hence when $\langle \Lambda \rangle/v$ falls below the turning point, the purchasing price turns from flat to rapidly rising. However, the turning point is sharp only in the limit of vanishing $v$. For finite values of $v$, the change is smoother. The inset of Fig. 1(b) shows that the capacity elasticity of price, $-vd\phi_p/d\langle \Lambda \rangle$, has a discontinuous slope at $\langle \Lambda \rangle/v = 0.622$, resembling a phase transition in many physical systems. Figure 3(A) shows the analytical result that prices rise rapidly when $\langle \Lambda \rangle/v$ falls below 0.622, accompanied by the disappearance of excess producers, and confirmed by simulation results in Fig. 3(B).

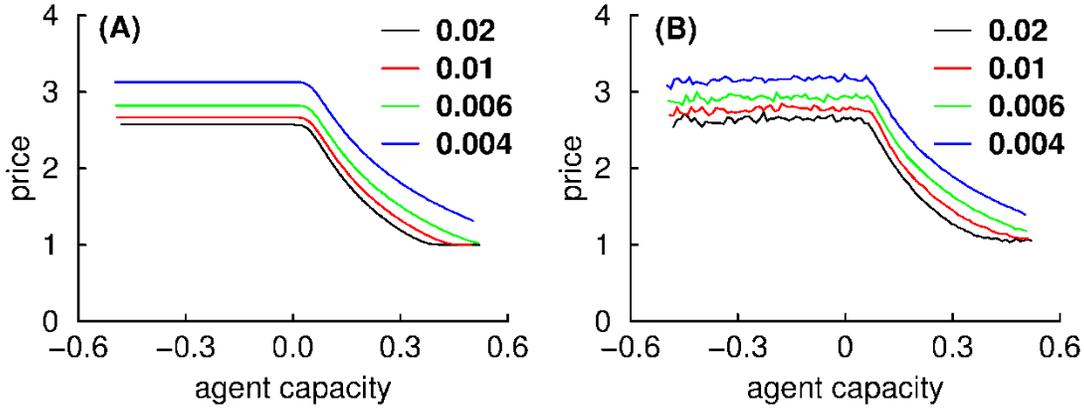

**Fig. 3. Prices rise rapidly when the average capacity falls below the turning point.** (A) The analytical result of the capacity dependence of the prices at different inventory levels at $v = 0.01$. (B) The corresponding simulation results.

When $\langle \Lambda \rangle$ falls further below $0.171v$, and the price rises to 3.67, even the quasi producers disappear (turquoise dot in Fig. 1(B)). However, since the excess resources held by the quasi producers are of the order $v$, the effect on the price behavior is much less pronounced. In this regime, the price rises with decreasing $\langle \Lambda \rangle$ asymptotically as $0.496v/\langle \Lambda \rangle$, and diverges when $\langle \Lambda \rangle/v$ approaches 0. Figure 3(A) shows the analytical result of the capacity dependence of the prices at different inventory levels. The prices rise rapidly when $\langle \Lambda \rangle/v$ falls below 0.622. Simulation results in Fig. 3(B) confirm the trend. The results also have an excellent agreement with those obtained by solving the Nash equilibrium equations (6).

In practice, most commodities are only considered essential when prices are not too high or stocks are sufficient. When prices become too high or stocks too low, the market will no longer consider the commodity essential. Buyers may switch to alternative commodities or at least refrain from purchasing. In the SM we consider a scenario whereby sales price reaches a maximum when the capital of agents is finite. We term this point the *yield point*.

**Comparison with Data**

To compare with commodities data, a common parameter to measure resource availability in commodity markets is the stocks-to-use ratio (SUR) defined as the amount of carryover stock of a commodity at the end of a period (usually a year) divided by the consumption during the same period [17]. While conventionally SUR is expressed as a percentage, it has the dimension of time, representing the duration in which stocks will be consumed by the market (assuming that no other resources are available). SUR is an important predictive tool of commodity prices [18]. For example, there is a strong negative correlation between cotton prices and SUR [19]. Similar trends were also observed in wheat and corn prices [5].

In general, a plot of the price of a commodity as a function of the SUR appears as a collection of scattered points, although a rough trend is often visible. One factor is that the data is gathered over many years or even decades, such that the data is interfered by many other factors, for instance changes in market needs. Here, we propose that the quality of data can be improved by defining the SUR elasticity of price,

$$E_p = -\frac{\text{change in price}}{\text{change in SUR}}. \tag{9}$$

In practice, we calculate the yearly elasticity of the commodities, and sort the corresponding SUR in order. Approximately 10 data points with consecutive SUR values are clustered for regression, and the slope of the cluster is taken to be the elasticity corresponding to the value of SUR averaged over the cluster. To compare with the theoretical prediction, the rescaling of the elasticity and SUR are explained in SM and plotted. Remarkably, a much clearer picture often emerges from this analysis.

For agricultural products in the U.S. market [20], Figs. 4(A)-(C) show the SUR dependence of the elasticity for long-grain rice, short-grain rice, cotton, and soybeans. The elasticities of these commodities have the common feature that they increase with decreasing SUR. We have also considered the data of other agricultural products. However, commodities such as honey and peanuts do not exhibit the behavior predicted by the trading model. This may be an indication that they are not essential and market demand would shrink if prices are too high.

Figure 4(D) is the composite plot of the four agricultural products illustrating their universal behavior. The plot is consistent with the prediction of our model showing that the elasticity increases with decreasing SUR. It is noted that the data can probably be fitted also be fitted with curves that continuously decrease with increasing SUR, such as in [5]. To provide a perspective on this point, it may be argued that the world economy has adjusted itself to the state of a low level of SUR, such that spare capacity is converted to other more efficient and profitable use of resources, rendering the turning point unobservable.

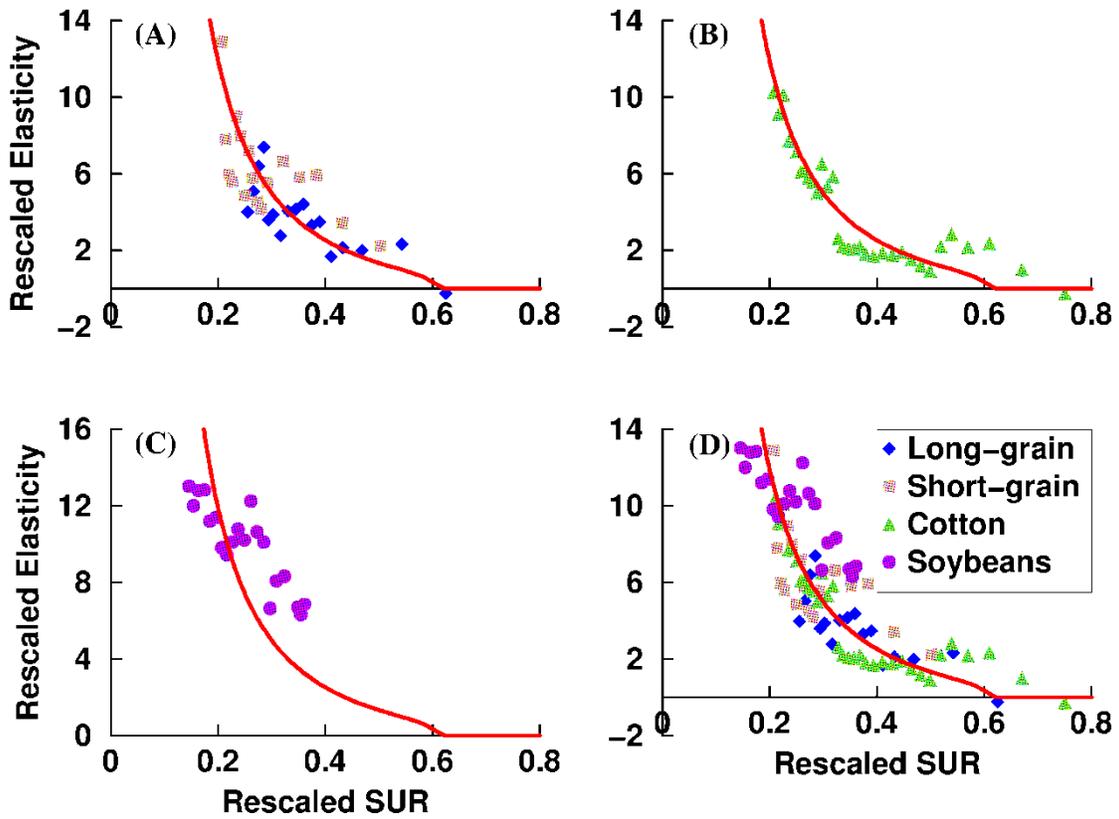

**Fig. 4. Elasticity of agricultural products.** The rescaled elasticity versus the rescaled SUR for (A) long-grain rice and short-grain rice 1983-2011, (B) cotton 1965-2010, (C) soybeans 1980-2012. The elasticities and the SURs are respectively rescaled by (A) 4.72 \$@1998/cwt/y and 0.385 y for long-grain rice, and 1.86 \$@1998/cwt/y and 0.830 y for short-grain rice, (B) 14.7 cents@1998/lb/y and 0.906 y, (C) 2.18 \$@1998/bu/y and 0.474 y. Each plotted point comes from a regression of 13 data points. (D) The composite plot of the four agricultural products. Solid curves: the pricing model.

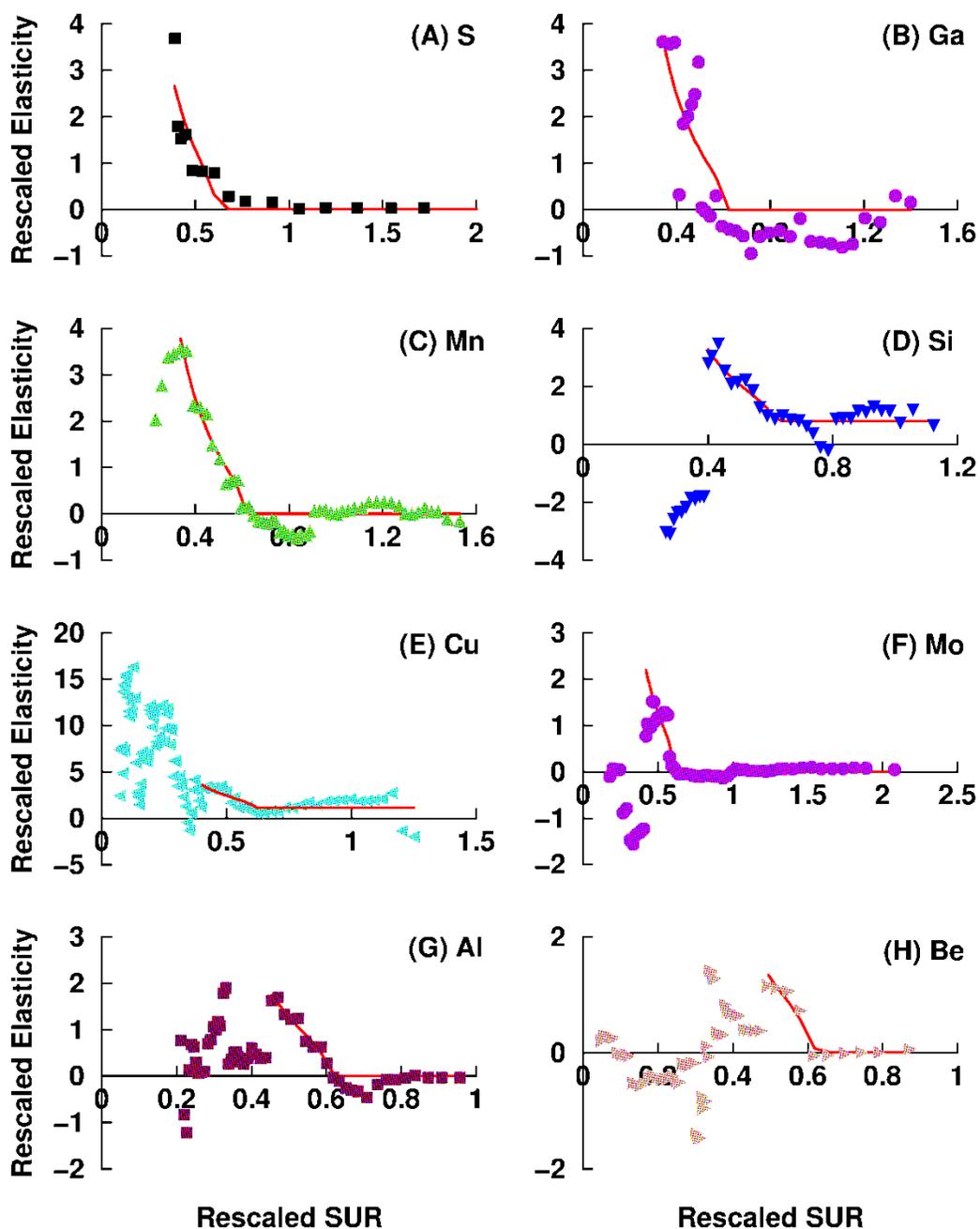

**Fig. 5. Elasticity of metals.** The rescaled elasticity versus the rescaled SUR for (A) sulfur 1928-2015, (B) gallium 1971-2015, (C) manganese 1949-2015, (D) silicon 1964-2015, (E) copper 1900-2015, (F) molybdenum 1941-2015, (G) aluminum 1949-2015, (H) beryllium 1941-2015. The elasticities and the SURs are respectively rescaled by (A) 2,900 $@1998/t/y and 0.0387 y, (B) 632,705 $@1998/t/y and 0.156 y, (C) 128 $@1998/t/y and 3.633 y, (D) 2,801 $@1998/t/y and 0.174 y, (E) 497 $@1998/t/y and 0.719 y, (F) 43,694 $@1998/t/y and 0.810 y, (G) 1,070 $@1998/t/y and 1.333 y, (H) 152,406 $@1998/t/y and 2.012 y. Each plotted point comes from a regression of 13 data points. Solid curves: the pricing model.

For metal prices [21], Figs. 5(A)-(B) show that the elasticities of sulfur and gallium exhibit a flat regime when SUR is high and rise when SUR falls below a point, agreeing with our model in the existence of turning points. On the other hand, Figs. 5(C)-(H) show that the elasticities of some other metals also has a flat and rising regime, but they increase only up to a yield point as SUR decreases. Below that point the elasticity either decreases with decreasing SUR or becomes broadly distributed, and even negative elasticities are observed.

The data at the turning points and yield points of these metals are shown in Table 1. It is interesting to note that despite the wide range of commodities, the SURs of most commodities at the turning points typically lie in the range 0.1 to 2.3 y, and their SURs at the yield points typically lie in the range 0.07 to 1.2 y. The SURs of gallium and sulfur are below 0.1 y at the turning points and the yield points are not observable from the data. To understand the typical value of the yield elasticity, we introduce the relative yield elasticity, defined as the elasticity at the yield point divided by the typical price of the commodity per unit of SUR at the yield point. The typical price of the commodity is calculated to be the average price in the elastic regime (between the turning and yield points). We find that the relative yield elasticity is in the range 0.1 to 1. It is plausible that the market mechanism determining the price of these commodities is rather universal.

We have also considered the data of other metals such as lead and nickel. They do not admit the behavior predicted by the pricing model, indicating that their prices may be affected by factors other than supply and demand.

| Commodity | SUR at turning point (year) | SUR at yield point (year) | Relative yield elasticity |
|---|---|---|---|
| Aluminum | 0.830 | 0.605 | 0.488 |
| Beryllium | 1.252 | 0.995 | 0.289 |
| Copper | 0.447 | 0.290 | 0.154 |
| Manganese | 2.261 | 1.223 | 1.053 |
| Molybdenum | 0.504 | 0.339 | 0.500 |
| Silicon | 0.108 | 0.070 | 0.402 |

Table 1. SUR at the turning and yield points and the relative yield elasticity for six metals.

In summary, the plots reveal an increasing elasticity with decreasing SUR before reaching the yield point if any. There exist turning points and yield points, separating the price behaviors into inelastic, elastic and yielded regimes. Most agricultural commodities predominantly cover the elastic regime. Data of other commodities such as cereal [22], crude oil [23, 24] and carbon trading [25, 26] have similar behaviors and are presented in the SM. On the other hand, less essential commodities such as metals have both turning points and yield points.

**Conclusion**

To understand how market structure drives commodity prices, we have proposed an agent-based model in which agents set their prices to maximize profits or minimize costs. The model is analytically solvable in a fully connected network, and the market behavior depends on only two mean-field parameters: the purchasing price $\phi_p$ and the demand coefficient $y$. The price behaviors in the different regimes can be attributed to the following ingredients of the model. (1) Resource is unevenly distributed and bounded among the agents. (2) Agents determine their prices through interactions, causing the market to self-organize into three types of agents depending on their capacities. Among them, the excess producers have excess resources and set their prices at the intrinsic value of the commodity. They act as a buffer for price stability. (3) The inclusion of inventory of individual agents smears out the sharp divergence of price to a turning point from the inelastic regime to the elastic one when the average capacity becomes comparable to the inventory level. This is in contrast to the model with no inventory, where the price jumps discontinuously to infinity when the systemwide resource vanishes. The transition at the turning point is due to a change in the market structure when excess resources are exhausted, and the market loses the buffering provided by excess producers. This mechanism is reminiscent of the Lewisian turning point, which describes the rise in wages of unskilled labor in developing economies when the labor market starts to run out of unskilled labor [7]. When the average resource is reduced further, the commodities become too expensive and the price rise may be restrained below the yield point.

To verify the model, we have analyzed the price history of agricultural products, metals and other commodities. We found that: (1) Elasticity versus SUR plots are much more interpretable than price versus SUR. This is probably because elasticity is based on short-term price changes, whereas a good plot of price versus SUR requires the long-term independence of the environment. (2) The elasticity versus SUR plot reveals two critical points: turning point and yield point. Three regimes are identified on decreasing SUR: inelastic, elastic and yielded. (3) Different data types have different characteristics. Only non-essential commodities have yield points. Most agricultural commodities cover the elastic regime only. Yielded regimes are present in most metals. (4) The data support the insights gained from our model.

We note that the price trends and the existence of the turning point are insensitive to details of the proposed model. Purchasing fractions other than the exponential function in Eq. (2) can be used, as long as they short-ranged functions of the price. Capacity distributions, other than the rectangular one, also exhibit similar behaviors as long as they have an upper bound. Similar predictions are also applicable to networks whose nodes have high but finite connectivity.

Since the commodities considered are storable, a natural extension of this study is to consider additional market demand for future consumption, and additional market supply from storing excess past production. Agents with this awareness tend to "buy low, sell high", thus providing an arbitrage mechanism on the commodity price. Indeed, a demand curve as a function of price was proposed with an additional component due to this storage demand when the price falls below a threshold [4]. Since the present study mainly pertains to the case of tight market supply and high prices, the effects of storage demand are neglected. Moreover, since storage introduces another variable to the model, the flexibility of the model may obscure the insight gained while offering limited benefit. A possible extension of this study is to consider the situation where the market supply fluctuates strongly in time, and agents adapt to the environment by adopting storage policies [3]. This problem can be approached by considering agents undertaking reinforcement learning [27], constituting an interesting future study.

The pricing model can be applied to prediction of prices in commodity markets. The wealth of generic results we have described in fully-connected networks can be generalized to other network structures after appropriate modifications to study specific features. Considering the finite connections of trading partners, we may consider networks with low connectivity. In such networks, the Onsager reactions of the pricing decisions will become significant, and it will be interesting to explore the effects of connectivity on the price and profit distributions. Another case of interest is geographical networks, in which interactions with neighbors depend on geographical distance, and regional dependence of prices and profits is likely to emerge. Extending the model to scale-free networks [28], we will be able to study the role played by hubs and satellites in distributing resources and gaining profits. By coupling the commodity network with another layer of network such as the financial network or the futures market network, we will be able to adopt the inter-dependent network approach [29] to consider broader issues of market stability.

**Materials and Methods**

**Design of the Simulation Experiment**

The clearing and price adjustment process of this trading model without inventory can be simulated in the following way: Each time step is divided into a large number of updating cycles. At the beginning of each time step, the position of each node, i.e., its net surplus (demand), and the transaction cost $E_i(t)$ of node $i$ are initialized to 0.

**Initialization:** At the beginning of each updating cycle, the position of node $i$ is increased (decreased) by $\Lambda_i$. The total number of purchasing orders during that cycle is the total number of negative positions of the consumers after rounding to the nearest integer.

**Purchasing orders:** A purchasing order is randomly chosen, and the buyer $i$ initiating the purchasing order is noted. A seller $j$ is selected with probability $r_{ij}$.

**Trading:** If the position of node $j$ is positive after rounding, the transaction is successful and the positions of $i$ and $j$ are raised and lowered by 1 respectively. Otherwise, node $j$ will act as a trader and randomly selects seller k with the probability $r_{jk}$. (In our formulation, node $i$ is not excluded.) If the position of node $k$ is positive after rounding, then the transactions between $i$ and $j$ and between $j$ and $k$ are both successful and the positions of $i$ and $k$ are updated. Otherwise, node $k$ will act as a trader and continues to search for a potential supplier until the chain of transactions can be carried out. In the rest of the cycle, other buyers are randomly chosen until the positions of all nodes become non-negative after rounding. The cycles are repeated so that the transaction costs can be better estimated. At the end of the time step, the price $\phi_i(t+1)$ is updated according to the current correlation between cost and price $\phi_i(t + 1) = \phi_i(t) - \eta w_i(t)\mathrm{sgn}\{[\phi_i(t) - \phi_i(t - 1)][E_i(t) - E_i(t - 1)]\}$, where $\eta$ is the adjustment rate, and $w_i(t)$ is a random number between 0 and 1 representing the stochastic nature of the pricing process.

To simulate the case of non-vanishing inventory level, the simulation is done analogously, except that: (1) the demands $d_i(t)$, dynamical dummy variables akin to the demands $\xi_i$, are initialized to 0 at the beginning of each time step; (2) the position of node $i$ is incremented by $\Lambda_i - v\ln\{1 + \exp[-|d_i(t)|/v]\}$ at the beginning of each updating cycle, reflecting its demands and inventory requirements; (3) the total number of purchasing orders during an updating cycle is the minimum of two numbers: the total number of negative positions of the consumers after rounding, or the total number of positive positions of the suppliers after rounding; and (4) the demand is updated according to $d_i(t+1) = d_i(t) + \eta\left[\left\langle n_i^{\mathrm{buy}}(t)\right\rangle - \Lambda_i - d_i(t)\right]$, where $\eta$ is the adjustment rate, and $\langle n_i^{\text{buy}}(t) \rangle$ is the average number of purchasing orders received by node *i* per updating cycle (irrespective of whether the transaction is successful or not), reflecting the change in demand given the activities of the previous time step.

**Statistical Analysis**

Simulations in Figs. 2(A-B) were done with $N = 100$, 1,000 samples and $\eta = 0.01$. Both distributions were obtained after 10,000 time steps and 200 updating cycles per time step. The capacities of the 100 agents were evenly spaced from $\langle \Lambda \rangle - 0.5$ to $\langle \Lambda \rangle + 0.5$. The bin sizes of the prices and costs are both 0.02.

The standard deviations of both distributions for a given value of the capacity change with the capacity. For the price distribution in the consumer regime, it decreases with the capacity from 0.25 to 0.15. In the balanced regime, it decreases with the capacity from 0.15 to 0.05 and then rises to 0.1. In the excess producer regime, it stays around 0.1. For the cost distribution in the consumer regime, it decreases with the capacity from 0.016 to 0.004. In the balanced regime, it increases with the capacity from 0.004 to 0.012. In the excess producer regime, it stays around 0.012.

Simulations in Fig. 3(B) were done with $N = 100$, 50 samples, $\eta = 0.01$, 10,000 time steps (the first 5,000 steps were used for equilibration) and 200 updating cycles per time step. The capacities of the 100 agents were evenly spaced from $\langle \Lambda \rangle - 0.5$ to $\langle \Lambda \rangle + 0.5$. The standard error of the mean prices remains below 0.04.

**Acknowledgments**

We thank Jack Raymond, Jacky Lam and Paul Choi for meaningful discussions, and Marvin Cohen for encouragement. This work is partially supported by grants from Research Grants Council of Hong Kong (grant numbers 604512, 605813 and 16322616) and the Leverhulme Trust RPG-2013-48.

**References**

[1] P. Perron, The Great Crash, The Oil Price Shock and The Unit Root Hypothesis, Econometrica **57**, 1361-1401 (1989).

[2] M. Lagi, K. Z. Bertrand, and Y. Bar Yam, The Food Crises and Political Instability in North Africa and Middle East, arXiv:1108.2455 (2011).

[3] R. L. Gustafson, Carryover Levels for Grains: A Method for Determining Amounts that are Optimal under Spectified Conditions, USDA Tech. Bulletin (United States Department of Agriculture) 1178 (1958).

[4] B. D. Wright, The Economics of Grain Price Volatility, Applied Economic Perspectives and Policy **33**, 32-58 (2011).

[5] P. C. Westscott and L. A. Hoffman, Price Determination for Corn and Wheat: The Role of Market Factors and Government Programs, USDA Tech. Bulletin (United States Department of Agriculture) 1878 (1999).

[6] L. A. Winters and D. Sapford, Primary Commodity Prices: Economic Models and Policy, Cambridge University Press, Cambridge MA (1990).

[7] W. A. Lewis, Economic Development with Unlimited Supplies of Labor, Manchester School of Economic and Social Studies **22**, 139-91 (1954).

[8] X. Zhang, J. Yang, and S. Wang, China has reached the Lewis turning point, China Econ. Rev. **22**, 542-554 (2011).

[9] M. Jackson, Social and Economic Networks, Princeton University Press, NJ (2008).

[10] M. Corominas-Bosch, Bargaining in a Network of Buyers and Sellers, J. Economic Theory **115**, 35-77 (2004).

[11] S. M. Kakade et al. Economic Properties of Social Networks, Advances in Neural Information Processing Systems **17**, 633-640, MIT Press, Cambridge, MA (2005).

[12] L. Blume, D. Easley, J. Kleinberg, E. Tardos, Trading Networks with Price-setting Agents, Games and Economic Behavior **67**, 36-50 (2009).

[13] S. M. Kakade, M. Kearns, and L. E. Ortiz, Graphical Economics, Learning Theory, Lecture Notes in Computer Science **3120**, 17-32 (2004).

[14] M. Mézard and A. Montanari, Information, Physics, and Computation, Oxford University Press, Oxford (2009).

[15] H. E. Stanley, Introduction to Phase Transitions and Critical Phenomena, Oxford University Press, Oxford (1971).

[16] P. Crompton and I. M. Xiarchos, Metal Prices and the Supply of Storage, Commodity Modeling and Pricing, P. V. Schaeffer (ed.), Wiley, NJ, 103-117 (2008).

[17] J. Womach, Agriculture: a Glossary of Terms, Programs, and Laws (Report for Congress) (2005).

[18] E. Bobenrieth et al, Stocks-to-use Ratios and Prices as Indicators of Vulnerability to Spikes in Global Cereal Markets, Agricultural Economics **44**, 43-52 (2013).

[19] Cotton Incorporated, Framing the Cotton Pricing Discussion, Textile Asia **41**, 51–52 (2010).

[20] U.S. Department of Agriculture Economics, Statistics, and Market Information System, http://usda.mannlib.cornell.edu/ (accessed 2017).

[21] T. D. Kelly and G.R. Matos, Historical Statistics for Mineral and Material Commodities in the United States (2015 version): U.S. Geological Survey Data Series 140, at http://minerals.usgs.gov/minerals/pubs/historical-statistics/ (2015).

[22] Food and Agriculture Organization of the United Nations, Global Information and Early Warning System, http://www.fao.org/giews/english/fo/ (2017).

[23] Organization of the Petroleum Exporting Countries, Monthly Oil Market Report (2001-2016).

[24] West Texas Research Group, Oil Price History and Analysis (2011).

[25] The EU Emissions Trading System, http://ec.europa.eu/clima/policies/ets/index\_en.htm

[26] Bloomberg stock code PNXCSPT2.

[27] R. S. Sutton and A. G. Barto, Reinforcement Learning: An Introduction, Second Edition, MIT Press, Cambridge, Massachusetts (2012).

[28] R. Albert and A. L. Barabási, Statistical Mechanics of Complex Networks, Reviews of Modern Physics **74**, 47-97 (2002).

[29] S. V. Buldyrev, R. Parshani, G. Paul, H. E. Stanley, and S. Havlin, Catastrophic Cascade of Failures in Interdependent Networks, Nature **464**, 1025-1028 (2010).

# Supplemental Materials of
# "How Market Structure Drives Commodity Prices"

Bin Li[1]†, K. Y. Michael Wong[1]*, Amos H. M. Chan[1]‡, Tsz Yan So[1]§,

Hermanni Heimonen[1]‖, Junyi Wei, and David Saad[2]

[1]Department of Physics, The Hong Kong University of Science and Technology, Hong Kong.
[2]The Nonlinearity and Complexity Research Group, Aston University, Birmingham B4 7ET, United Kingdom.

1. **Green's Function Approach to the Trading Model**

First, consider the cost function of node $i$,

$$E_i = \sum_{j \in \partial i} \xi_i r_{ij} \phi_j - \sum_{j \in \partial i} \xi_j r_{ji} \phi_i. \tag{S1}$$

determining the expenditure and revenue, respectively. When there is price adjustment $\delta\phi_i$, $E_i$ is changed by

$$\delta E_i = \sum_{j \in \partial i} \delta\xi_i r_{ij} \phi_j - \sum_{j \in \partial i} \delta\xi_j r_{ji} \phi_i - \sum_{j \in \partial i} \xi_j \left( r_{ji} \delta\phi_i + \phi_i \delta r_{ji} \right). \tag{S2}$$

Consequently, the demand of $i$ changes because its neighbors $j$ adjust their fractions of resource purchased from node $i$ due to $\delta\phi_i$. Furthermore, there are reaction effects due to the adjustment of the demands of the neighbors $j$, $\delta\xi_j$, and hence the change $\delta\xi_i$ is

$$\delta\xi_i = f'\left( \sum_{j \in \partial i} \xi_j r_{ji} - \Lambda_i \right)\left( \sum_{j \in \partial i} \delta\xi_j r_{ji} + \sum_{j \in \partial i} \xi_j \delta r_{ji} \right). \tag{S3}$$

In turn, the change of the demands $\delta\xi_j$ is given by a similar equation, except that there is no driving term due to the adjustment of the resource fractions,

$$\delta\xi_j = f'\left( \sum_{k \in \partial j} \xi_k r_{kj} - \Lambda_j \right)\left( \sum_{k \in \partial j \setminus i} \delta\xi_k r_{kj} + \delta\xi_i r_{ij} \right). \tag{S4}$$

Note that the demands of $j$ involve those of nodes $k$ that are the second neighbors of $i$. The series of demand equations can be written, each time involving a new generation of neighbors one step further away. Hence, it is convenient to introduce Green's functions given by

$$\delta\xi_k = G_{kj} \delta\xi_j. \tag{S5}$$

Substituting into Eq. (S4), we have

$$\delta\xi_j = \frac{\delta\xi_i r_{ij} f_j'}{1 - \sum_{k \in \partial j \setminus i} G_{kj} r_{kj} f_j'}, \tag{S6}$$

where $f_j' \equiv f'\left( \sum_{k \in \partial j} \xi_k r_{kj} - \Lambda_j \right)$. This yields a recursion relation of the Green's functions given by

$$G_{ji} = \frac{\delta \xi_j}{\delta \xi_i} = \frac{r_{ij} f_j'}{1 - \sum_{k \in \partial j \setminus i} G_{kj} r_{kj} f_j'}. \tag{S7}$$

This expression can be generalized to any pair of neighbors. Furthermore, we can write the Green's function from $i$ to itself. From Eq. (S3),

$$\delta \xi_i = \frac{\sum_{j \in \partial i} \xi_j \delta r_{ji} f_i'}{1 - \sum_{j \in \partial i} G_{ji} r_{ji} f_i'}. \tag{S8}$$

Hence we can write

$$\delta \xi_i = G_i \sum_{j \in \partial i} \xi_j \delta r_{ji}, \tag{S9}$$

where $G_i$ is given by

$$G_i = \frac{f_i'}{1 - \sum_{j \in \partial i} G_{ji} r_{ji} f_i'}. \tag{S10}$$

To complete the description, we substitute the above expression back to Eq. (S2) to obtain the equation for the optimal price,

$$\delta E_0 = G_i \sum_{j \in \partial i} \xi_j \delta r_{ji} \left( \sum_{k \in \partial i} r_{ik} \phi_k - \sum_{k \in \partial i} G_{ki} r_{ki} \phi_i \right) - \sum_{j \in \partial i} \xi_j \left( r_{ji} \delta \phi_i + \phi_i \delta r_{ji} \right). \tag{S11}$$

By setting $\partial E_i / \partial \phi_i = 0$ and noting that $\delta r_{ji} = -r_{ji}(1 - r_{ji}) \delta \phi_i$, we arrive at an equation for the optimal price,

$$\phi_i = \frac{1}{1 + G_i \sum_{j \in \partial i} G_{ji} r_{ji}} \left( \frac{\sum_{j \in \partial i} \xi_j r_{ji}}{\sum_{j \in \partial i} \xi_j r_{ji}(1 - r_{ji})} + G_i \sum_{j \in \partial i} r_{ij} \phi_j \right). \tag{S12}$$

For networks with connectivity of the order $c$, $r_{ji} \sim c^{-1}$. Hence we can see from Eq. (S7) that $G_{ji} \sim c^{-1}$, and from Eq. (S10) that $G_i \approx f_i'$. We then obtain

$$\phi_i = \frac{\sum_{j \in \partial i} \xi_j r_{ji}}{\sum_{j \in \partial i} \xi_j r_{ji}(1 - r_{ji})} + f' \left( \sum_{j \in \partial i} \xi_j r_{ji} - \Lambda_i \right) \sum_{j \in \partial i} r_{ij} \phi_j. \tag{S13}$$

This is identical to Eq. (6) in the main text in the limit of a fully connected network, that is, the reaction effects embodied in the Green's functions $G_{ji}$ are negligible in this limit.

2. **Small $\nu$ Expansion**

In the full connectivity limit, Eq. (S13) can be further simplified. Since $r_{ji} \sim O(N^{-1})$, the first term on the right hand side reduces to 1. The second term depends on two macroscopic parameters. First, we have

$$\sum_{j\neq i}\xi_j r_{ji} = \frac{\sum_{j\neq i}\xi_j}{e^{-\phi_i}+\sum_{j\neq i}e^{-\phi_j}}e^{-\phi_i} \approx \frac{\langle\xi\rangle}{\langle e^{-\phi}\rangle}e^{-\phi_i} = ye^{-\phi_i}. \tag{S14}$$

Hence the *demand coefficient* is given by

$$y \equiv \frac{\langle\xi\rangle}{\langle e^{-\phi}\rangle}. \tag{S15}$$

Second, we have the average purchasing price given by

$$\phi_p \equiv \sum_{j\neq i} r_{ij}\phi_j = \sum_{j\neq i}\frac{e^{-\phi_j}\phi_j}{e^{-\phi_i}+\sum_{k\neq i}e^{-\phi_k}} \approx \frac{\langle\phi e^{-\phi}\rangle}{\langle e^{-\phi}\rangle}, \tag{S16}$$

This reduces $\phi_i$ in Eq. (S13) to the solution of

$$\phi = 1 + \phi_p f'(ye^{-\phi}-\Lambda). \tag{S17}$$

Since $f'(x) = [1+\exp(-x/v)]^{-1}$, it is convenient to invert the functional relationship between $\phi$ and $\Lambda$, yielding

$$\Lambda = ye^{-\phi} + v\ln\left(\frac{1+\phi_p-\phi}{\phi-1}\right). \tag{S18}$$

To derive expressions for $y$ and $\phi_p$, it is difficult to write the small $v$ expansion due to their implicit dependence of v. We turn to consider the explicit integrals to achieve this. For $\langle e^{-\phi}\rangle$, we write it as an area integral in the space of $\Lambda$ and $\phi$,

$$\langle e^{-\phi}\rangle = \int_a^b d\Lambda\rho(\Lambda)\int_{\phi(\Lambda)}^\infty d\phi e^{-\phi}. \tag{S19}$$

where $a$ and $b$ are the lower and upper bounds of the capacity distribution respectively. As shown in the Fig. S1, interchanging the order of integrating $\Lambda$ and $\phi$ yields

$$\langle e^{-\phi}\rangle = \int_{\phi(b)}^{\phi(a)} d\phi e^{-\phi}\int_{\Lambda(\phi)}^b d\Lambda\rho(\Lambda) + \int_{\phi(a)}^\infty d\phi e^{-\phi}\int_a^b d\Lambda\rho(\Lambda). \tag{S20}$$

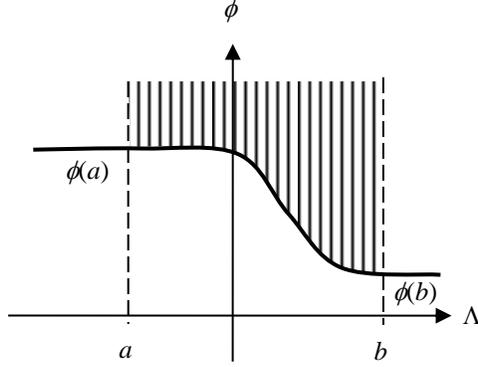

**Fig. S1.** The area of integration in Eqs. (S19) and (S20).

For the rectangular distribution with $a = \langle \Lambda \rangle - 1/2$ and $b = \langle \Lambda \rangle + 1/2$, we have

$$\langle e^{-\phi} \rangle = \int_{\phi(b)}^{\phi(a)} d\phi e^{-\phi} [b - \Lambda(\phi)] + e^{-\phi(a)}. \tag{S21}$$

In the small $v$ limit and when the average capacity $\langle \Lambda \rangle$ is of the order $v^0$, $\phi(a) \approx 1 + \phi_p$ and $\phi(b) \approx 1$. Hence, after integration, we have

$$\langle e^{-\phi} \rangle = be^{-1} - ae^{-1-\phi_p} - \frac{y}{2}e^{-2} + \frac{y}{2}e^{-2-2\phi_p} - v\int_1^{1+\phi_p} d\phi e^{-\phi} \ln\left(\frac{1+\phi_p-\phi}{\phi-1}\right). \tag{S22}$$

Similarly, we can write

$$\langle \phi e^{-\phi} \rangle = \int_{\phi(b)}^{\phi(a)} d\phi e^{-\phi}(\phi-1)[b-\Lambda(\phi)] + \phi(a)e^{-\phi(a)}. \tag{S23}$$

After integration,

$$\langle \phi e^{-\phi} \rangle = be^{-1} - a(1+\phi_p)e^{-1-\phi_p} - \frac{y}{4}e^{-2} + \frac{y}{4}(1+2\phi_p)e^{-2-2\phi_p}.$$
$$- v\int_1^{1+\phi_p} d\phi e^{-\phi}(\phi-1)\ln\left(\frac{1+\phi_p-\phi}{\phi-1}\right). \tag{S24}$$

For $\langle \xi \rangle$, we note that the expression appropriate for small $v$ expansion is

$$\xi = v\ln\left[1+\exp\left(\frac{ye^{-\phi(\Lambda)}-\Lambda}{v}\right)\right] = \max(ye^{-\phi(\Lambda)}-\Lambda,0) + v\ln\left[1+\exp\left(-\frac{|ye^{-\phi(\Lambda)}-\Lambda|}{v}\right)\right]. \tag{S25}$$

From Eq. (S18), $ye^{-\phi(\Lambda)} - \Lambda = 0$ when $\phi = 1 + \phi_p/2$. Thus, we have the integral expression

$$\langle \xi \rangle = \int_a^{y\exp(-1-\phi_p/2)} d\Lambda \rho(\Lambda)\left(ye^{-\phi(\Lambda)} - \Lambda\right)$$

$$+ \int_a^{y\exp(-1-\phi_p/2)} d\Lambda \rho(\Lambda) v \ln\left[1 + \exp\left(-\frac{ye^{-\phi(\Lambda)} - \Lambda}{v}\right)\right]$$

$$+ \int_{y\exp(-1-\phi_p/2)}^b d\Lambda \rho(\Lambda) v \ln\left[1 + \exp\left(\frac{ye^{-\phi(\Lambda)} - \Lambda}{v}\right)\right]. \tag{S26}$$

Substituting Eq. (S18), we have

$$\langle \xi \rangle = \int_a^{y\exp(-1-\phi_p/2)} d\Lambda \rho(\Lambda)\left(ye^{-\phi(\Lambda)} - \Lambda\right) + \int_a^{y\exp(-1-\phi_p/2)} d\Lambda \rho(\Lambda) v \ln\frac{\phi_p}{\phi(\Lambda) - 1}$$

$$+ \int_{y\exp(-1-\phi_p/2)}^b d\Lambda \rho(\Lambda) v \ln\frac{\phi_p}{1 + \phi_p - \phi(\Lambda)}. \tag{S27}$$

Expressing $\langle \xi \rangle$ as area integrals,

$$\langle \xi \rangle = \int_a^{y\exp(-1-\phi_p/2)} d\Lambda \rho(\Lambda) \int_{\phi(\Lambda)}^\infty d\phi\, y e^{-\phi} - \int_a^{y\exp(-1-\phi_p/2)} d\Lambda \rho(\Lambda) \Lambda$$

$$+ \int_a^{y\exp(-1-\phi_p/2)} d\Lambda \rho(\Lambda) \int_{\phi(\Lambda)}^{1+\phi_p} d\phi \frac{v}{\phi - 1} + \int_{y\exp(-1-\phi_p/2)}^b d\Lambda \rho(\Lambda) \int_1^{\phi(\Lambda)} d\phi \frac{v}{1 + \phi_p - \phi}. \tag{S28}$$

Interchanging the order of integrating $\Lambda$ and $\phi$ yields

$$\langle \xi \rangle = \int_{1+\phi_p/2}^{\phi(a)} d\phi\, y e^{-\phi} \int_{\Lambda(\phi)}^{y\exp(-1-\phi_p/2)} d\Lambda \rho(\Lambda) + \int_{\phi(a)}^\infty d\phi\, y e^{-\phi} \int_a^{y\exp(-1-\phi_p/2)} d\Lambda \rho(\Lambda)$$

$$- \int_a^{y\exp(-1-\phi_p/2)} d\Lambda \rho(\Lambda)\Lambda + \int_{1+\phi_p/2}^{\phi(a)} d\phi \frac{v}{\phi - 1} \int_{\Lambda(\phi)}^{y\exp(-1-\phi_p/2)} d\Lambda \rho(\Lambda)$$

$$+ \int_{\phi(a)}^{1+\phi_p} d\phi \frac{v}{\phi - 1} \int_a^{y\exp(-1-\phi_p/2)} d\Lambda \rho(\Lambda) + \int_{\phi(b)}^{1+\phi_p/2} d\phi \frac{v}{1 + \phi_p - \phi} \int_{y\exp(-1-\phi_p/2)}^{\Lambda(\phi)} d\Lambda \rho(\Lambda)$$

$$+ \int_1^{\phi(b)} d\phi \frac{v}{1 + \phi_p - \phi} \int_{y\exp(-1-\phi_p/2)}^b d\Lambda \rho(\Lambda). \tag{S29}$$

For the rectangular distribution we have

$$\langle \xi \rangle = \int_{1+\phi_p/2}^{\phi(a)} d\phi y e^{-\phi} \left[ y e^{-1-\phi_p/2} - \Lambda(\phi) \right] + \int_{\phi(a)}^{\infty} d\phi y e^{-\phi} \left[ y e^{-1-\phi_p/2} - a \right] - \frac{y^2}{2} e^{-2-\phi_p} + \frac{a^2}{2}$$

$$+ \int_{1+\phi_p/2}^{\phi(a)} d\phi \frac{v}{\phi-1} \left[ y e^{-1-\phi_p/2} - \Lambda(\phi) \right] + \int_{\phi(a)}^{1+\phi_p} d\phi \frac{v}{\phi-1} \left[ y e^{-1-\phi_p/2} - a \right]$$

$$+ \int_{\phi(b)}^{1+\phi_p/2} d\phi \frac{v}{1+\phi_p-\phi} \left[ \Lambda(\phi) - y e^{-1-\phi_p/2} \right] + \int_{1}^{\phi(b)} d\phi \frac{v}{1+\phi_p-\phi} \left[ b - y e^{-1-\phi_p/2} \right] \tag{S30}$$

In the small $v$ limit, the integrals ranging from $\phi(a)$ to $1 + \phi_p$ and from 1 to $\phi(b)$ are negligible. After integration, one obtains

$$\langle \xi \rangle = \frac{y^2}{2} e^{-2-2\phi_p} + \left( \frac{1}{2} - \langle \Lambda \rangle \right) y e^{-1-\phi_p} + \frac{1}{2} \left( \frac{1}{2} - \langle \Lambda \rangle \right)^2 - v \int_{1+\phi_p/2}^{1+\phi_p} d\phi y e^{-\phi} \ln\left( \frac{1+\phi_p-\phi}{\phi-1} \right)$$

$$+ v \int_{1+\phi_p/2}^{1+\phi_p} d\phi \left( \frac{y e^{-1-\phi_p/2} - y e^{-\phi}}{\phi-1} \right) + v \int_{1}^{1+\phi_p/2} d\phi \left( \frac{y e^{-\phi} - y e^{-1-\phi_p/2}}{1+\phi_p-\phi} \right). \tag{S31}$$

After obtaining the expressions $\langle e^{-\phi} \rangle$, $\langle \phi e^{-\phi} \rangle$ and $\langle \xi \rangle$ in Eqs. (S22), (S24) and (S31) respectively, we substitute them into the self-consistent equations for $\phi_p$ and $y$ in Eqs. (S16) and (S15), yielding

$$\left( \frac{1}{2} + \langle \Lambda \rangle - \frac{y}{2} e^{-1} \right) \phi_p e^{-1} = \left( \frac{1}{2} + \langle \Lambda \rangle \right) e^{-1} + \left( \frac{1}{2} - \langle \Lambda \rangle \right) e^{-1-\phi_p} - \frac{y}{4} e^{-2} + \frac{y}{4} e^{-2-2\phi_p}$$

$$+ v \int_{1}^{1+\phi_p} d\phi e^{-\phi} (1+\phi_p-\phi) \ln\left( \frac{1+\phi_p-\phi}{\phi-1} \right), \tag{S32}$$

and

$$\frac{1}{2} y^2 e^{-2} - \left( \frac{1}{2} + \langle \Lambda \rangle \right) y e^{-1} + \frac{1}{2} \left( \frac{1}{2} - \langle \Lambda \rangle \right)^2 + v \int_{1}^{1+\phi_p/2} d\phi y e^{-\phi} \ln\left( \frac{1+\phi_p-\phi}{\phi-1} \right)$$

$$+ v \int_{1+\phi_p/2}^{1+\phi_p} d\phi \left( \frac{y e^{-1-\phi_p/2} - y e^{-\phi}}{\phi-1} \right) + v \int_{1}^{1+\phi_p/2} d\phi \left( \frac{y e^{-\phi} - y e^{-1-\phi_p/2}}{1+\phi_p-\phi} \right) = 0. \tag{S33}$$

### 3. The $v = 0$ Limit

In the limit $v = 0$, Eqs. (S32) and (S33) reduce to

$$\phi_p = \frac{\frac{1}{2} + \langle \Lambda \rangle + \left(\frac{1}{2} - \langle \Lambda \rangle\right) e^{-\phi_p} - \frac{y}{4} e^{-1} + \frac{y}{4} e^{-1-2\phi_p}}{\frac{1}{2} + \langle \Lambda \rangle - \frac{y}{2} e^{-1}}, \tag{S34}$$

and

$$\frac{1}{2}\left(ye^{-1}\right)^2 - \left(\frac{1}{2} + \langle \Lambda \rangle\right) ye^{-1} + \frac{1}{2}\left(\frac{1}{2} - \langle \Lambda \rangle\right)^2 = 0. \tag{S35}$$

The solution of the quadratic equation for $ye^{-1}$ is

$$ye^{-1} = \left(\sqrt{\frac{1}{2}} - \sqrt{\langle \Lambda \rangle}\right)^2. \tag{S36}$$

When $\langle \Lambda \rangle$ approaches 1/2, all nodes of the network have non-negative capacities, and the demand coefficient is expected to vanish. Hence the other solution $ye^{-1} = \left(\sqrt{\frac{1}{2}} + \sqrt{\langle \Lambda \rangle}\right)^2$ is discarded.

Note that Eq. (S36) implies $ye^{-1} \leq 1/2 + \langle \Lambda \rangle$. This means that excess producers exist in the range implies $ye^{-1} \leq \Lambda \leq 1/2 + \langle \Lambda \rangle$. When $\langle \Lambda \rangle$ approaches 0, the lower bound of the capacities of the excess producers, distribution, $ye^{-1}$, approaches the upper bound of the capacity distribution, $1/2 + \langle \Lambda \rangle$. Hence the fraction of excess producers in the network approaches 0. However, for finite $\langle \Lambda \rangle$, excess producers always exist in the limit of $v = 0$.

Before the price diverges, it will approach a finite value $\phi_c$ given by Eq. (S34) at $\langle \Lambda \rangle = 0$,

$$\phi_c = \frac{3}{2} + 2e^{-\phi_c} + \frac{1}{2} e^{-2\phi_c}. \tag{S37}$$

This yields $\phi_c = 1.83$ with an infinite slope at $\langle \Lambda \rangle = 0$, as shown in Fig. S2. The corresponding demand coefficient is $y_c = e/2$. When $\langle \Lambda \rangle$ falls below 0, the price diverges discontinuously. However, as we shall see, this discontinuity is absent when we consider the behavior at small values.

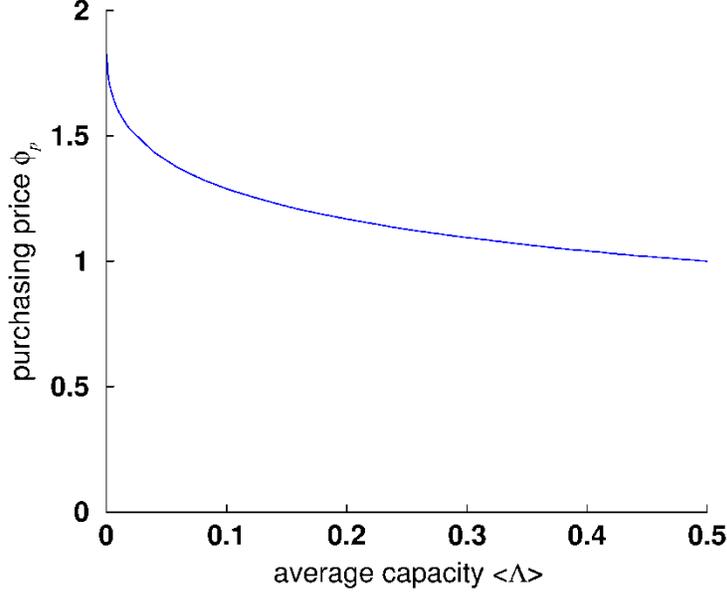

**Fig. S2. Price without inventory.** The dependence of the purchasing price on the average capacity in the $v = 0$ limit.

## 4. The Small $v$ Limit

When $v$ is finite but small, correction terms in Eq. (S33) have to be included, yielding the quadratic equation

$$\frac{1}{2}(ye^{-1})^2 - \left(\frac{1}{2} + \langle \Lambda \rangle\right)ye^{-1} + \frac{1}{2}\left(\frac{1}{2} - \langle \Lambda \rangle\right)^2 + Jv = 0, \tag{S38}$$

where

$$J = y \int_1^{1+\phi_p/2} d\phi \left[\frac{e^{-\phi} - e^{-2-\phi_p+\phi}}{1+\phi_{\text{sale}} - \phi} + e^{-\phi} \ln\left(\frac{1+\phi_p - \phi}{\phi - 1}\right)\right]. \tag{S39}$$

The solution of Eq. (S38) is given by

$$ye^{-1} = b - \sqrt{2\langle \Lambda \rangle - 2Jv}. \tag{S40}$$

Solution exists when $\langle \Lambda \rangle \geq Jv$. No solution will imply that there are no excess producers in the network. Hence, the critical capacity for the existence of excess producers is given by

$$\frac{\Lambda_c}{v} = y \int_1^{1+\phi_c/2} d\phi \left[\frac{e^{-\phi} - e^{-2-\phi_c+\phi}}{1+\phi_c - \phi} + e^{-\phi} \ln\left(\frac{1+\phi_c - \phi}{\phi - 1}\right)\right]. \tag{S41}$$

The numerical result is $\Lambda_c/v = 0.622$.

## 5. The Regime of No Excess Producers

In this regime, the expressions $\langle e^{-\phi}\rangle$, $\langle e^{-\phi}\phi\rangle$ and $\langle\xi\rangle$ have to be modified. In particular, in the limit of small $v$, $\phi(b)$ is no longer equal to 1. Instead, we have $\phi(b) \approx \ln(y/b)$. Hence, the expression for $\langle e^{-\phi}\rangle$ is replaced by

$$\langle e^{-\phi}\rangle = \int_{\ln(y/b)}^{1+\phi_p} d\phi e^{-\phi}[b-\Lambda(\phi)] + e^{-1-\phi_p}. \tag{S42}$$

After integration in the limit of small $\langle\Lambda\rangle$,

$$\langle e^{-\phi}\rangle = \frac{1}{2y}\left(\frac{1}{2}+\langle\Lambda\rangle\right)^2 + \left(\frac{1}{2}-\langle\Lambda\rangle\right)e^{-1-\phi_p} + \frac{y}{2}e^{-2-2\phi_p} - v\int_{\ln(2y)}^{1+\phi_p} d\phi e^{-\phi} \ln\left(\frac{1+\phi_p-\phi}{\phi-1}\right). \tag{S43}$$

Similarly for $\langle\phi e^{-\phi}\rangle$, Eq. (S24) is replaced by

$$\langle\phi e^{-\phi}\rangle = \int_{\ln(y/b)}^{1+\phi_p} d\phi e^{-\phi}(\phi-1)[b-\Lambda(\phi)] + (1+\phi_p)e^{-1-\phi_p}. \tag{S44}$$

After integration,

$$\langle\phi e^{-\phi}\rangle = \frac{b^2}{4y} + \frac{b^2}{2y}\ln\left(\frac{y}{b}\right) + \left(\frac{1}{2}-\langle\Lambda\rangle\right)(1+\phi_p)e^{-1-\phi_p} + \frac{y}{4}(1+2\phi_p)e^{-2-2\phi_p}$$
$$- v\int_{\ln(2y)}^{1+\phi_p} d\phi e^{-\phi}(\phi-1)\ln\left(\frac{1+\phi_p-\phi}{\phi-1}\right). \tag{S45}$$

For $\langle\xi\rangle$, Eq. (S31) is replaced by

$$\langle\xi\rangle = \int_{1+\phi_p/2}^{1+\phi_p} d\phi y e^{-\phi}[y e^{-1-\phi_p/2}-\Lambda(\phi)] + \int_{1+\phi_p}^{\infty} d\phi y e^{-\phi}[y e^{-1-\phi_p/2}-a] - \frac{y^2}{2}e^{-2-\phi_p} + \frac{a^2}{2}$$
$$+ v\int_{1+\phi_p/2}^{1+\phi_p} d\phi\left(\frac{y e^{-1-\phi_p/2}-\Lambda(\phi)}{\phi-1}\right)$$
$$+ v\int_{\ln(2y)}^{1+\phi_p/2} d\phi\left(\frac{\Lambda(\phi)-y e^{-1-\phi_p/2}}{1+\phi_p-\phi}\right) + v\int_{1}^{\ln(2y)} d\phi\left(\frac{1/2-y e^{-1-\phi_p/2}}{1+\phi_p-\phi}\right). \tag{S46}$$

Noting that only terms of order $v$ are modified, the result is

$$\langle\xi\rangle = \frac{1}{2}\left(ye^{-1-\phi_p}\frac{1}{2}-\langle\Lambda\rangle\right)^2$$
$$-v\int_{1+\phi_p/2}^{1+\phi_p}d\phi ye^{-\phi}\ln\left(\frac{1+\phi_p-\phi}{\phi-1}\right)+v\int_{1+\phi_p/2}^{1+\phi_p}d\phi\left(\frac{ye^{-1-\phi_p/2}-\Lambda(\phi)}{\phi-1}\right)$$
$$+v\int_{\ln(2y)}^{1+\phi_p/2}d\phi\left(\frac{\Lambda(\phi)-ye^{-1-\phi_p/2}}{1+\phi_p-\phi}\right)+v\int_1^{\ln(2y)}d\phi\left(\frac{1/2-ye^{-1-\phi_p/2}}{1+\phi_p-\phi}\right). \tag{S47}$$

Hence, in the regime of no excess producers, we have to use the replaced Eqs. (S43), (S45) and (S47) for expressing $\langle e^{-\phi}\rangle$, $\langle\phi e^{-\phi}\rangle$ and $\langle\xi\rangle$ respectively, and the self-consistent equations (S32) and (S34) are thus replaced by

$$\frac{1}{2y}\left(\frac{1}{2}+\langle\Lambda\rangle\right)^2\left[\phi_p-\frac{1}{2}-\ln\left(\frac{y}{b}\right)\right]=\left(\frac{1}{2}-\langle\Lambda\rangle\right)e^{-1-\phi_p}+\frac{y}{4}e^{-2-2\phi_p}$$
$$+v\int_{\ln(2y)}^{1+\phi_p}d\phi e^{-\phi}(1+\phi_p-\phi)\ln\left(\frac{1+\phi_p-\phi}{\phi-1}\right), \tag{S48}$$

and

$$\frac{\langle\Lambda\rangle}{v}=y\int_{\ln(2y)}^{1+\phi_p/2}d\phi\left[\frac{e^{-\phi}-e^{-2-\phi_p+\phi}}{1+\phi_p-\phi}+e^{-\phi}\ln\left(\frac{1+\phi_p-\phi}{\phi-1}\right)\right]+y\int_1^{\ln(2y)}d\phi\left(\frac{1/2y-e^{-2-\phi_p+\phi}}{1+\phi_p-\phi}\right). \tag{S49}$$

In the limit of small $v$ and $\langle\Lambda\rangle \sim v$, Eq. (S48) reduces to

$$\phi_p = \frac{1}{2}+\ln 2y+4ye^{-1-\phi_p}+2y^2e^{-2-2\phi_p}. \tag{S50}$$

Let $x = 1 + \phi_p - \ln(2y)$. Then $x$ satisfies Eq. (S37). This implies

$$y = \frac{1}{2}e^{1+\phi_p-\phi_c}. \tag{S51}$$

Hence for a given price $\phi_p$, we can find the corresponding value of $y$. Substituting $\phi_p$ and $y$ into Eq. (S49), we can obtain the corresponding value of $\langle\Lambda\rangle/v$ directly.

## 6. The Regime of No Quasi Producers

Even in the regime of no excess producers, there are quasi producers whose resource lies between $ye^{-1-\phi_p/2}$ and $b$, and whose price is set between $\ln(y/b)$ and $1 + \phi_p/2$. Their capacities are higher than their out-currents by order $v$. When $\langle\Lambda\rangle/v$ drops sufficiently, $ye^{-1-\phi_p/2}$ approaches $1/2$ and even the quasi producers disappear. Substituting this condition into Eq. (S51), we find

$$\phi_{c2} = 2\phi_c. \tag{S52}$$

This transition takes place when the price becomes 3.67, $\langle\Lambda\rangle/v$ becomes 0.171, and $y$ becomes 4.25.

When $\langle\Lambda\rangle/v$ falls below 0.171, $\langle e^{-\phi}\rangle$ and $\langle\phi e^{-\phi}\rangle$ still follow Eqs. (S43) and (S5), but $\langle\xi\rangle$ in Eq. (S47) is replaced by

$$\langle\xi\rangle = \int_{\ln(y/b)}^{1+\phi_p} d\phi ye^{-\phi}[b-\Lambda(\phi)] + \int_{1+\phi_p}^{\infty} d\phi ye^{-\phi}\left[-\frac{b^2}{2}+\frac{a^2}{2}\right] + v\int_{\ln(y/b)}^{1+\phi_p} d\phi\left(\frac{b-\Lambda(\phi)}{\phi-1}\right). \tag{S53}$$

After integration,

$$\langle\xi\rangle = \frac{1}{2}\left(\frac{1}{2}-\langle\Lambda\rangle\right)^2 + \left(\frac{1}{2}-\langle\Lambda\rangle\right)ye^{-1-\phi_p} + \frac{y^2}{2}e^{-2-2\phi_p}$$
$$+ v\int_{\ln(2y)}^{1+\phi_p} d\phi\left(\frac{1/2-ye^{-\phi}}{\phi-1}\right) - v\int_{\ln(2y)}^{1+\phi_p} d\phi ye^{-\phi}\ln\left(\frac{1+\phi_p-\phi}{\phi-1}\right). \tag{S54}$$

Equation (S49) is replaced by

$$\frac{\langle\Lambda\rangle}{v} = \int_{\ln(2y)}^{1+\phi_p} d\phi\left(\frac{1/2-ye^{-\phi}}{\phi-1}\right). \tag{S55}$$

Noting that the lower integration limit is equal to $1 + \phi_s - \phi_c$, we find in the limit of $\langle\Lambda\rangle/v$ approaching 0, the price diverges as

$$\phi_{\text{sale}} = \frac{(\phi_c - 1 + e^{-\phi_c})}{2(\langle\Lambda\rangle/v)} = \frac{0.496}{\langle\Lambda\rangle/v}. \tag{S56}$$

## 7. A Possible Mechanism for the Yield Point

So far we have assumed that the commodities are essential. However, when prices are too high, some agents cannot afford to purchase all the commodities they need. This provides one possible mechanism for the occurrence of the yield point. To analyze the situation, we consider the case that the capital of the consumers is finite and equal to $C_0$. Due to their limited capital, a fraction of consumers with the largest consumption will purchase less commodities and their capacity will be effectively replaced by a more positive capacity $\Lambda_0$.

These agents purchase commodities of the amount $ye^{-1-\phi_p} - \Lambda_0$ at the price $\phi_p$, and sell commodities of the amount $ye^{-1-\phi_p}$ at the price $1 + \phi_p$. Hence $\Lambda_0$ is given by

$$C_0 = (ye^{-1-\phi_p} - \Lambda_0)\phi_p - ye^{-1-\phi_p}(1+\phi_p), \tag{S57}$$

implying

$$\Lambda_0 = -\frac{C_0 + ye^{-1-\phi_p}}{\phi_p}. \tag{S58}$$

The self-consistent equations for $\phi_p$ and $y$ can then be derived, using the capacity distribution in which those capacities lying between $a$ and $\Lambda_0$ are now replaced by $\Lambda_0$. Assume that the yield point lies in the regime of no excess producers. Then the expressions for $\langle e^{-\phi} \rangle$ and $\langle \phi e^{-\phi} \rangle$ remain identical to Eqs. (S43) and (S45) respectively. On the other hand, the equation for $\langle \xi \rangle$ becomes

$$\langle \xi \rangle = \frac{1}{2}\left( ye^{-1-\phi_p}\frac{1}{2} - \langle \Lambda \rangle \right)^2 - \frac{1}{2}\left( \Lambda_0 + \frac{1}{2} - \langle \Lambda \rangle \right)^2$$
$$-v\int_{1+\phi_p/2}^{1+\phi_p} d\phi ye^{-\phi}\ln\left(\frac{1+\phi_p-\phi}{\phi-1}\right) + v\int_{1+\phi_p/2}^{1+\phi_p} d\phi\left(\frac{ye^{-1-\phi_p/2} - \Lambda(\phi)}{\phi-1}\right)$$
$$+v\int_{\ln(2y)}^{1+\phi_p/2} d\phi\left(\frac{\Lambda(\phi) - ye^{-1-\phi_p/2}}{1+\phi_p-\phi}\right) + v\int_{1}^{\ln(2y)} d\phi\left(\frac{1/2 - ye^{-1-\phi_p/2}}{1+\phi_p-\phi}\right). \tag{S59}$$

The self-consistent equation for $\phi_p$ remains the same as Eq. (S51), whereas the self-consistent equation for $y$ becomes

$$\frac{\langle \Lambda \rangle}{v} = y\int_{\ln(2y)}^{1+\phi_p/2} d\phi\left[\frac{e^{-\phi} - e^{-2-\phi_p+\phi}}{1+\phi_p-\phi} + e^{-\phi}\ln\left(\frac{1+\phi_p-\phi}{\phi-1}\right)\right]$$
$$+y\int_{1}^{\ln(2y)} d\phi\left(\frac{1/2y - e^{-2-\phi_p+\phi}}{1+\phi_p-\phi}\right) - \frac{(\Lambda_0 + 1/2 - \langle \Lambda \rangle)^2}{2v}. \tag{S60}$$

This equation differs from Eq. (S49) only in the last term, showing that when $\Lambda_0 \sim -1/2 + O(\sqrt{v})$, there will be a significant change in the solution. This change takes place when $y$ and $\phi_p$ are related via Eq. (S58) by

$$y = \left(\frac{\phi_p}{2} - C_0\right) e^{1+\phi_p}. \tag{S61}$$

Eliminating $y$ from Eq. (S51),

$$\phi_p = 2C_0 + e^{-\phi_c}. \tag{S61}$$

Hence when the price reaches this level, the price will be pinned when the average capacity continues to decrease according to Eq. (S60). The elasticity $-d\phi_p/d\langle\Lambda\rangle$ will drop discontinuously to 0.

## 8. Elasticity Curves of Agricultural Products, Metals and Other Commodities

In this section we first describe how the rescaled elasticity curves in Figs. 4, 5 and S3 are constructed. Then we analyze other commodities not mentioned in the main text such as cereal, crude oil and carbon trading.

A. Elasticity Curves

Elasticity curves in Figs. 4 and S3 are constructed by first performing a linear regression of the data points. The $x$ intercept of the regression line is then taken to be turning point, thus fixing the SUR scale. We then rescale the elasticities by the elasticity scale, and define the cost function to be the mean square difference between the rescaled empirical elasticities and their theoretical predicted values. The cost function is minimized with respect to the elasticity scale.

Elasticity curves in Figs. 5(A)-(H) are constructed with additional procedures since the data points are noisier and the elasticity curves need to be fitted to the three regimes (inelastic, elastic and yield). First, we propose that the elasticity of a commodity is the sum of its value in the elastic regime (as described by the curve in Fig. 1(B) inset) and a non-negative background elasticity $E_{\text{back}}$. As shown in Fig. 1(B), $\phi_p$ approaches an effectively constant value (= 1.83) when $\langle\Lambda\rangle \gg v$, but as indicated in Fig. S2 of SM, $\phi_p$ continues to drops gently when $\langle\Lambda\rangle$ increases to the order of $v^0$. This gentle drop may give rise to the background elasticity. We then define a cost function that is proportional to the mean square difference between the rescaled empirical elasticities and their theoretical predicted values. Only those data points considered to be in the inelastic and elastic regimes are included in the cost function, which is minimized with respect to the SUR scale $S_{\text{SUR}}$, the elasticity scale $S_E$ and $E_{\text{back}}$. To prevent the overflow of the elasticity scale or the underflow of SUR scale, the proportionality constant in the cost function is $S_E^n / S_{\text{SUR}}$, where $n = 2$ for all metals (except for sulfur where $n = 1$).

To determine the range of data points in the inelastic and elastic regimes, we successively expand the range into the yielded regime. When the range does not include data points in the yielded regime, the cost function per data point is roughly constant. However, when the range starts to cover the yielded regime, the cost function per data point starts to increase sharply, or may even diverge. This provides a clear-cut criterion to demarcate the yielded regime.

### A. Cereal

Cereal data are available from the UN FAO yearly food outlooks [22]. These reports provided global annual average prices and major exporters' SUR for different commodities, such as wheat, coarse grains and rice. Although data from 1992 to 2012 are available, the data show a discontinuity in the SUR from 1995 to 1996. According to the report from February 2001, the discontinuity was due to significant data changes when the cereal stocks estimates in China (Mainland) were revised [22]. Hence we focus on wheat and coarse grains data from 1991 to 2010 excluding 1995 for the elasticity versus SUR plot. As shown in Figs. S3(A)-(B), both wheat and coarse grains data follow the trend predicted by our model in the elastic regime.

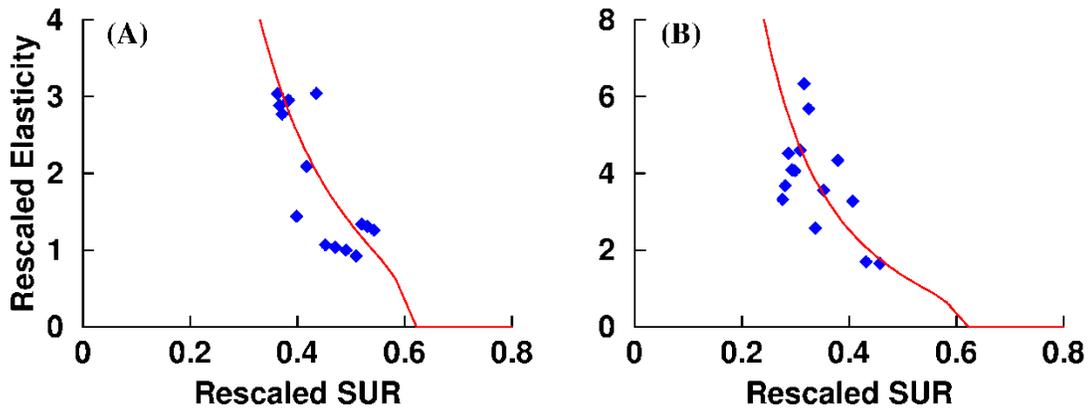

**Fig. S3. Elasticity of cereal.** The rescaled elasticity versus the rescaled SUR for (A) wheat, (B) coarse grains, both from 1991 to 2012 and excluding 1995. To enable comparison with the trading model (the solid curve with a turning point at 0.622), the elasticities and SURs are rescaled by (A) 448 US$@1998/tonne/y and 0.684 y, (B) 223 US$@1998/tonne/y and 0.563 y. Each plotted point comes from a regression of 7 data points.

### B. Crude Oil

We collect SUR and price data of crude oil from OPEC Monthly Oil Market Reports [23]. Since oil usage has seasonal variations, we consider data averaged over four consecutive quarters. The plot of the price is shown in the inset of Fig. S4(A). It shows the general trend of increasing price on decreasing SUR. By plotting the elasticity in Fig. S4(A), we find that the elasticity is positive and rather constant. This shows that besides the market factors considered in our model, there may be additional factors such as the efforts of OPEC countries in trying to control prices by regulating production quota [24].

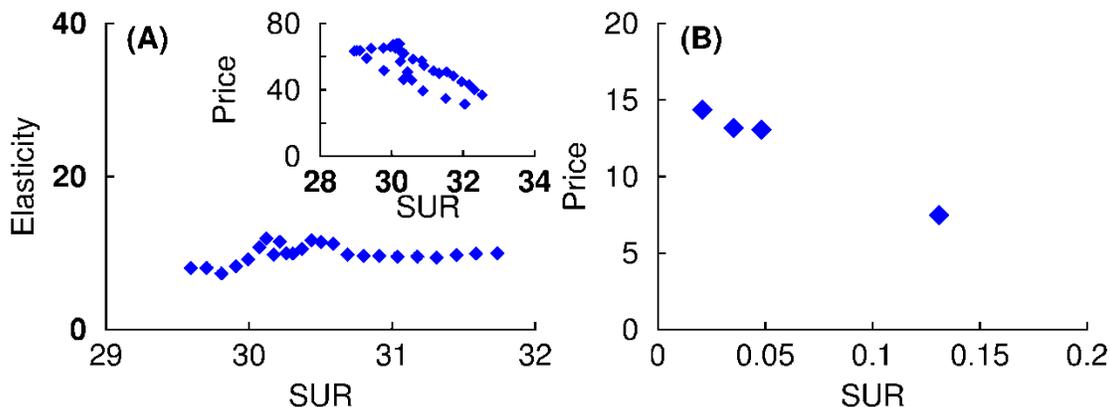

**Fig. S4. Elasticity of crude oil and carbon trading.** (A) The elasticity (in US$@Jun01/barrel/mbpd) versus the SUR of crude oil from 1st quarter of 2007 to 2nd quarter of 2015. 1 mbpd is equal to 1 million barrels per day. Each data point is based on the average of the considered quarter and three subsequent quarters. Each plotted point represents a regression of 11 data points. Inset: Basket crude oil prices in US\$@Jun01 per barrel versus SUR. (B) The price of carbon permits in Euros per tonne of carbon dioxide versus SUR from 2009 to 2012.

### C. Carbon Trading

We further studied carbon trading in the European Union Emission Trading System [25]. EU-wide carbon permit prices can be obtained from the French stock exchange up to 2012 [26]. Daily prices are averaged annually. The SUR of carbon trading is defined as the EU-wide carbon emission allocation minus the actual release, divided by the actual release. In contrast with other physical commodities, negative SURs are allowed for carbon trading, but penalty was imposed on non-compliance. In practice, a negative SUR was only found for the years 2008 when carbon trading entered phase 2. Figure S4(B) shows that the price decreases with increasing SUR. Since data points are too few, we have not attempted the elasticity plot.